\documentclass[aps,prl,twocolumn,amsmath,amssymb,nofootinbib,showpacs,superscriptaddress]{revtex4-1}
\usepackage[english]{babel}
\usepackage{latexsym}
\usepackage{graphics}
\usepackage{epsfig}
\usepackage{color}
\usepackage{bm}
\usepackage{amsmath}
\usepackage{amssymb}

\begin{document}

\title{Stable dilute supersolid of two-dimensional dipolar bosons}

\author{Zhen-Kai Lu}
\affiliation{Max-Planck-Institut f\"ur Quantenoptik,
Hans-Kopfermann-Stra{\ss}e 1, 85748 Garching, Germany}

\author{Yun Li}
\affiliation{Centre for Quantum and Optical Science, Swinburne
University of Technology, Melbourne, Victoria, 3122, Australia}

\author{D.~S.~Petrov}
\affiliation{Universit\'e Paris-Sud, CNRS, LPTMS, UMR8626, Orsay,
F-91405, France}

\author{G.~V.~Shlyapnikov}
\affiliation{Universit\'e Paris-Sud, CNRS, LPTMS, UMR8626, Orsay,
F-91405, France} \affiliation{\mbox{Van der Waals-Zeeman Institute,
University of Amsterdam, Science Park 904, 1098 XH Amsterdam, The
Netherlands}} \affiliation{Russian Quantum Center, Novaya street
100, Skolkovo, Moscow region 143025, Russia} \affiliation{Wuhan
Institute of Physics and Mathematics, Chinese Academy of Sciences,
Wuhan 430071, China}

\date{\today}

\begin{abstract}

We consider two-dimensional bosonic dipoles oriented perpendicularly
to the plane. On top of the usual two-body contact and long-range
dipolar interactions we add a contact three-body repulsion as
expected, in particular, for dipoles in the bilayer geometry with
tunneling. The three-body repulsion is crucial for stabilizing the
system, and we show that our model allows for stable continuous
space supersolid states in the dilute regime and calculate the
zero-temperature phase diagram.
\end{abstract}

 \maketitle

Recent advances in the field of cold polar molecules \cite{Ni,Carr}
and magnetic atoms \cite{Lev,Ferlaino} interacting via long-range
dipole-dipole forces make it realistic to create novel many-body
quantum states in these systems. For polar molecules, ultracold
chemical reactions observed at JILA \cite{Jin,Jin2} and leading to a
rapid decay of the system can be suppressed by tightly confining the
molecules to a (quasi)two-dimensional (2D) geometry, orienting the
dipoles perpendicularly to the plane of their translational motion,
and thus inducing a strong intermolecular repulsion
\cite{Bohn1,Baranov1,Ye}. Therefore, 2D geometries are intensively
discussed in the context of ultracold dipolar gases
\cite{Baranov2008,Baranov-Dalmonte}, together with possible
experiments with non-reactive molecules, such as
NaK\cite{Zwierlein1,Zwierlein2} and RbCs\cite{Grimm, Cornish}.

The studies of ultracold dipolar gases may open perspectives for the
observation of supersolidity. This remarkable quantum phenomenon
combines superfluidity with a crystalline order
\cite{Gross1,Kirzhnits} (see \cite{prokofiev} for review). It is
still under debate as to what extent experimental results in solid
helium prove the existence of this conceptually important phase
\cite{Balibar2010}. On the other hand, supersolidity is rather well
understood theoretically for soft-core two-body potentials
\cite{prokofiev,note3,Rica,Henkel,Cinti,Saccanti} which can be
realized, for example, in Rydberg-dressed atomic gases. However,
such supersolids require a dense regime with at least several
particles within the interaction range, which can be difficult to
achieve. The same holds for supersolids discussed for 2D dipolar
Bose gases \cite{Kurbakov} near the gas-solid phase transition
\cite{Buchler2007,Astrakharchik2007}. It is thus an open question
whether supersolids can exist in the dilute regime. The creation of
such supersolids, especially if they are tunable regarding the
lattice period, will allow for studies of non-conventional
superfluid properties of supersolids and other aspects of
supersolidity. Dilute 2D dipolar bosons may show the (helium-like)
roton-maxon structure of the spectrum by fine-tuning the short-range
part of the interaction potential and can be made unstable with
respect to periodic modulations of the order parameter (roton
instability) \cite{Gora2}. However, instead of forming a supersolid
state when approaching such an instability, the gas collapses
\cite{Gora3,Cooper2}.

In this Letter we predict a stable supersolid state in a dilute
two-dimensional dipolar   system. In contrast to the earlier
proposed soft-core supersolids, where the lattice period is of the
order of the core radius, in our case it is tunable by varying the
density and the dipole moment. In addition to the contact two-body
term ($g_2$) and the dipole-dipole long-range tail characterized by
the dipole moment $d$, we include a contact repulsive three-body
term ($g_3$) which may prevent the collapse. Three-body forces are
ubiquitous and arise naturally in effective field theories when one
integrates out some of  the high-energy degrees of freedom in the
system \cite{Hammer}. In particular, our model can be realized for
dipoles in the bilayer geometry with interlayer tunneling
\cite{Petrov}. Tracing out the degree of freedom associated with the
layer index one obtains an effective single-layer model in which
$g_2$ and $g_3$ can be independently controlled by tuning the
interlayer tunneling amplitude. Here we work out the phase diagram
of this model and identify stable  uniform and supersolid states.

The Hamiltonian of the system reads
\begin{eqnarray}
\mathcal{H}&=&-\int d^2r\hat\psi^{\dag}({\bf
r})\frac{\hbar^2\nabla^2}{2m}\hat\psi({\bf r})
+\mathcal{H}_2   \nonumber \\
&+&\frac{g_3}{6}\int d^2r\hat\psi^{\dag}({\bf
r})\hat\psi^{\dag}({\bf r})\hat\psi^{\dag}({\bf r})\hat\psi({\bf
r})\hat\psi({\bf r})\hat\psi({\bf r}), \label{H}
\end{eqnarray}
where $\hat\psi({\bf r})$ is the bosonic field operator, $m$ is the
particle mass, and the normalization volume is set equal to unity.
The first term in Eq.~(\ref{H}) corresponds to the kinetic energy,
the third one to the contact three-body repulsion ($g_3>0$), and the
two-body interaction Hamiltonian $\mathcal{H}_2$ at low energies can
be substituted by an effective momentum-dependent (pseudo)potential
(see, e.g., \cite{Ali})
\begin{equation} \label{V}
\tilde V({\bf k},{\bf k'})=\tilde V(|{\bf k}-{\bf k'}|)=  g_2-2\pi
d^2|{\bf k}-{\bf k'}|,
\end{equation}
where ${\bf k}$ and ${\bf k'}$ are the incoming and outgoing
relative momenta, $g_2$ is the contact term which depends on the
short-range details of the two-body potential, and the
momentum-dependent part corresponds to the long-range dipolar tail
for dipoles oriented perpendicularly to the plane of their
translational motion. We thus have
\begin{equation}\label{H2}
\!\mathcal{H}_2\!=\!\frac{1}{2}\!\!\int\!\!\!d^2rd^2r'\hat\psi^{\dag}\!({\bf
r})\hat\psi^{\dag}\!({\bf r'})\!\!\sum_{{\bf q}}\!\tilde
V(q)e^{i{\bf q}({\bf r}'-{\bf r})}\hat\psi({\bf r})\hat\psi({\bf
r'}).\!\!\!
\end{equation}

The onset of supersolidity is frequently associated with the
presence of a low-lying roton minimum in the excitation spectrum
\cite{Pitaevskii,Nozieres,Rica}. In our case the standard Bogoliubov
approach for a uniform Bose condensate of density $n$ gives the
excitation spectrum
\begin{equation}  \label{epsilon}
\epsilon(k)=\sqrt{E_k^2+2E_k(g_2n+g_3n^2-2\pi nd^2k)},
\end{equation}
where $E_k=\hbar^2k^2/2m$, and we assume that $(g_2+g_3n)>0$. The
validity conditions for the mean-field approach read
\begin{equation}  \label{ineq}
nr_*^2\ll 1;\,\,\,\,\,\,\,\,m(g_2+g_3n)/\hbar^2\ll 1,
\end{equation}
where $r_*=md^2/\hbar^2$ is a characteristic range  of the
dipole-dipole interaction. The structure of the spectrum is
characterized by a dimensionless parameter $\beta$ given by
\begin{equation}   \label{beta}
\beta=\gamma/(1+g_2/g_3n);\,\,\,\,\,\,\gamma=4\pi^2\hbar^2r_*^2/mg_3.
\end{equation}
The excitation energy $\epsilon(k)$ shows a roton-maxon  structure
(local maximum and minimum at finite $k$) for $\beta$ in the
interval $8/9<\beta<1$, and at $\beta=1$ the roton minimum touches
zero. For $\beta>1$ the excitation energies become imaginary, and
the uniform superfluid (U) is dynamically unstable and is no longer
the ground state.

A promising candidate for the new ground state is a supersolid state
in which the condensate wavefunction is a superposition of a
constant term and a lattice-type function of coordinates
\cite{Gross1,Kirzhnits,Pitaevskii,Rica}. We considered various
lattice structures and found that the ground state can be either a
triangular lattice supersolid (T) or a stripe supersolid (S)
\cite{SM}.  For T, the lattice is built up on three vectors in the
$x,y$ plane of the translational motion, with the angle of $2\pi/3$
between each pair: ${\bf k_1}=(k, 0)$, ${\bf k_2}=(-k/2,
\sqrt{3}k/2)$, and ${\bf k_3}=(-k/2, -\sqrt{3}k/2)$, while for the S
phase the density modulation depends only on one wave vector ${\bf
k}=(k, 0)$.

The variational ansatz for the condensate wavefunction of the T
phase then takes the form:
\begin{equation}  \label{superpsi1}
\psi_{T}({\bf r})=\sqrt{n}\left(\cos\theta+\sqrt{2/3} \sin\theta
e^{i\Phi}\sum_{i}\cos{\bf k}_i{\bf r}\right),
\end{equation}
and for the S phase we have:
\begin{equation}     \label{psistripe}
\psi_{S}({\bf r}) =\sqrt{n}\left(\cos\theta +\sqrt{2}\sin\theta
e^{i\Phi}\cos kx\right),
\end{equation}
which satisfies the normalization condition $\int d{\bf r}
|\psi_{T(S)}({\bf r})|^2 =n$, with $n$ being the mean density. The
variational parameters of the wavefunctions are $\theta$, $\Phi$,
and $k$. Density modulations appear at $\theta\neq 0$, and thus
$\theta$ is the order parameter which exhibits the U to  supersolid
transition. We have checked that the lowest energy always
corresponds to $\Phi=0$ and for brevity we omit this parameter.

For obtaining the energy functionals of the T and S states, we
replace the field operators in Eqs.~(\ref{H}) and (\ref{H2}) with
$\psi_{T}({\bf r})$ and with $\psi_{S}({\bf r})$, respectively. This
yields
\begin{equation}  \label{Epsi1}
\!\!\mathcal{E}_i\!=\!\left[E_kn\!-\!4\pi n^{2} d^{2}k
\mathcal{D}_i(\theta)\!\right]\!\sin^{\!2}\theta\!+\!g_2n^2
\mathcal{C}_i(\theta)\!+\!g_3n^3\mathcal{T}_i(\theta),\!\!\!
\end{equation}
where the symbol $i$ stands for $T$ and $S$, and the functions
$\mathcal{D}_{T(S)}(\theta)$, $\mathcal{C}_{T(S)}(\theta)$, and
$\mathcal{T}_{T(S)}(\theta)$ are related to the two-body
dipole-dipole, two-body contact, and three-body contact
interactions, respectively \cite{SM}.

By minimizing Eq.~(\ref{Epsi1}) with respect to $k$ we obtain
\begin{equation} \label{Epsi2}
\mathcal{E}_i(\!k_{mi})=g_2n^2 \mathcal{C}_i(\theta)\!+g_3
n^3(\mathcal{T}_i(\theta)-2\gamma\sin^2\!\theta \,
{\mathcal{D}_i^2(\theta})),\!
\end{equation}
where $k_{mi}\!\!=\!\!4\pi nr_*\mathcal{D}_i(\theta)$. In the dilute
limit of Eq.(\ref{ineq}) the particle number per unit modulation
volume is $n(2\pi/k_{mi})^2\!\!\sim\!\!1/nr_*^2\!\!\gg\!\!1$, which
justifies the mean-field approach.

The energy functional $\mathcal{E}_{T(S)}$ can be expanded in powers
of $\theta$. The zero-order term $\mathcal{E}(\theta=0)=g_2
n^2/2+g_3n^3/6$ gives the energy density of the uniform state. The
expansion of $\mathcal{E}_{T}$ contains terms $\propto \theta^3$
\cite{SM}, which is a consequence of the fact that the vectors ${\bf
k}_1$, ${\bf k}_2$, and ${\bf k}_3$ form a closed triangle
(``triad'', ${\bf k}_1+{\bf k}_2+{\bf k}_3=0$) \cite{Kirzhnits}.  In
contrast, the expansion of  $\mathcal{E}_{S}$ contains only even
powers of $\theta$. According to the Ginzburg-Landau theory
\cite{Landau, Binder}, the U-supersolid transition should occur to
the T phase and it is expected to be first order, so that $\theta$
jumps from 0 to a finite value.  However, deeply in the supersolid
regime the states with different structures are energetically
competing and, in particular, the stripe phase can become the ground
state of the system.

First-order transitions are convenient to analyse in the
grand-canonical picture. We obtain the phase diagram by
variationally minimizing  the grand potential
$\Omega=\mathcal{E}_{T(S)}-\mu n$ with respect to $\theta$ and $n$
for given values of the chemical potential $\mu$ and the interaction
parameters $g_2$, $g_3$ and $d$. We have checked the phase diagram
by employing the full numerical minimization of the grand potential
density, which is equivalent to solving the corresponding
Gross-Pitaevskii (GP) equation \cite{SM}.

First, let us consider $g_2=0$. In this case the energy functional
$\mathcal{E}$  only contains terms $\propto n^3$, and the phase
diagram is determined by a single dimensionless parameter $\gamma$
defined in Eq.~(\ref{beta}). The U to T transition occurs before the
roton minimum touches zero (for $g_2=0$ we have $\beta=\gamma$),
namely at $\gamma_0\simeq 0.99$, where $\theta$ jumps from 0 to
$0.0946$. The inverse compressibility
$\kappa^{-1}=\partial\mu/\partial n=6\mathcal{E}/n^2$ is positive
for $\gamma$ smaller than approximately 1.4, indicating the
existence of a stable supersolid state. However, our numerics
predicts the collapse instability at about $\gamma_c\approx 0.88$
and indicates that for lower values of $\gamma$ the ground state is
a uniform superfluid. The discrepancy between the numerics and
variational ansatz comes from the fact that the latter does not take
into account higher order momentum harmonics.

For $g_2\neq 0$, we turn to the rescaled dimensionless density
$\tilde n=ng_3/|g_2|$, chemical potential $\tilde\mu=\mu g_3/g_2^2$,
and grand potential $\tilde\Omega_{T(S)} =(g_3^2/|g_2|^3)
\Omega_{T(S)} =\mathcal{\tilde E}_{T(S)}-\tilde\mu\tilde n$. The
rescaled energy functional is given by
\begin{equation}
\mathcal{\tilde E}_i=[\mathcal{T}_i(\theta) -2\gamma\sin^2 \theta
\mathcal{D}_i^2(\theta)]\tilde n^3+{\rm sgn} (g_2) \tilde n^2
\mathcal{C}_i(\theta). \label{Erescaled}
\end{equation}
The phase diagram can be presented in the parameter space
($\tilde\mu,\gamma)$ and the phases are characterized by
$\theta\in[-\pi/2,\pi/2]$ and $\tilde n$. One can easily see that in
the high-density regime $\tilde\Omega_{T(S)}$ is dominated by the
term $[\mathcal{T}_{T(S)}(\theta)-2\gamma\sin^2\!\theta
\mathcal{D}_{T(S)}^2(\theta)]\tilde n^3$, whereas the two-body
contact interaction, i.e., the term containing
$\mathcal{C}_{T(S)}(\theta)$, becomes irrelevant. In this case the T
phase has a lower $\tilde\Omega$ than the S phase, and we obtain the
same stability condition as in the case of $g_2=0$. Numerically we
find that the phase diagram for $g_2>0$ contains only a stable U
state at $\gamma<\gamma_c$ and the region of collapse for
$\gamma>\gamma_c$.

\begin{figure}
\includegraphics[width=1.\hsize,clip]{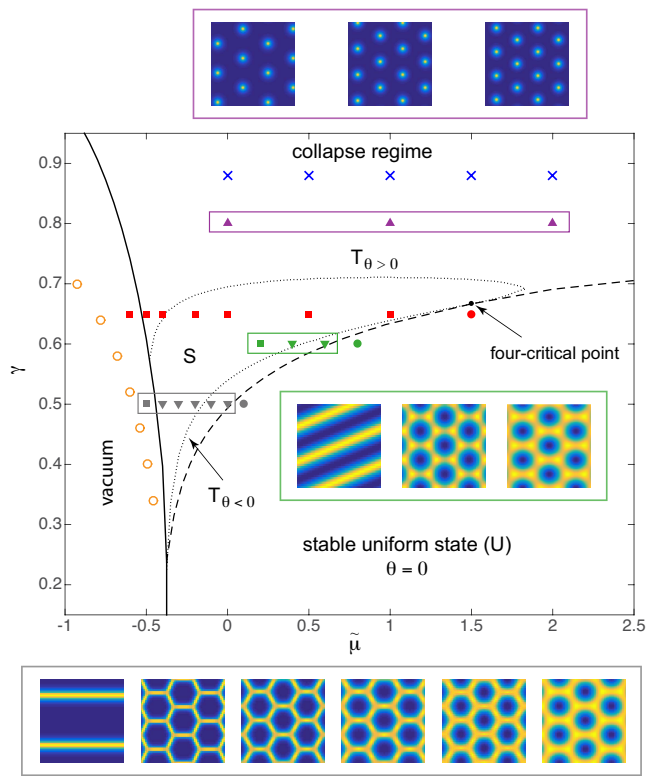}
\caption{(color online) Phase diagram for $g_2<0$. Continuous curves
correspond to transitions between different phases obtained from the
variational ans\"atze (\ref{superpsi1}) and (\ref{psistripe}).
Increasing $\gamma$ one passes the U-T transition (dashed), then T-S
(lower dotted curve), and S-T (upper dotted curve). To the left of
the solid black curve the ground state of the system is vacuum. The
black dot is the four-critical point for the U, S, and two T phases.
The symbols indicate our numerical results: the filled circles are
inside the U phase, the downward and upward pointing triangles are
inside the T${}_{\theta<0}$ and T${}_{\theta>0}$ phases,
respectively, and squares are in the stripe phase. The empty circles
are on the vacuum-stripe line and crosses are at the collapse
instability border. The color-coded pictures show density profiles
corresponding to the symbols in the phase diagram put in frames: the
upper set (violet frame) contains three points of the
T${}_{\theta>0}$ phase at $\gamma=0.8$, the middle set (green frame)
shows one point in the S phase and two points in the hexagonal
T${}_{\theta<0}$ phase at $\gamma=0.6$, and the lower set (grey
frame) corresponds to the six points at $\gamma=0.5$.}
\label{NegaPhase}
\end{figure}

The situation is quite different for $g_2<0$. The phase diagram is
shown in Fig.~\ref{NegaPhase} where all continuous curves correspond
to the variational results and all symbols to the exact numerical
solution of the GP equation. Let us first discuss the variational
results. The dashed curves mark the U-T${}_{\theta<0}$ and
U-T${}_{\theta>0}$ transitions, which occur for $\tilde\mu <3/2$ and
$\tilde\mu>3/2$, respectively. These are first order transitions
which weaken on approaching the point $\tilde\mu=3/2$, $\gamma=2/3$
(black dot). The same holds for the dotted curves, which correspond
to the transitions from the T phases to the S phase. The black dot
thus stands as a four-critical point and it is the only place in the
phase diagram where the transitions are second order and occur when
the roton minimum touches zero. In this case the grand potential
$\tilde\Omega={\rm const} + O(\theta^4)$, i.e., the terms $\propto
\theta^2$ and $\propto \theta^3$ are absent.

The region on the left of the black solid curve in
Fig.~\ref{NegaPhase} is the vacuum state: $\tilde n=0$, $\Omega=0$.
Directly on the curve, vacuum can coexist with matter which has a
finite density and zero pressure. We thus are dealing with a
self-trapped droplet state \cite{Bulgac}.  With increasing $\gamma$,
the vacuum curve eventually bends towards negative $\tilde\mu$ and
tends to the variational collapse line $\gamma\approx 1.4$ (not
shown).

By solving the GP equation numerically we observe that the overall
structure of the phase diagram is well captured by the variational
ans\"atze (\ref{superpsi1}) and (\ref{psistripe}). Close to the
four-critical point the agreement is quantitative, which is
generally expected in the regions where $\theta\ll 1$. Far from this
point we see that the exact collapse line moves to $\gamma\approx
0.88$ (crosses in Fig.~\ref{NegaPhase}) and the vacuum curve (empty
orange circles) bends towards negative $\tilde\mu$ faster than its
variational version. The rest of the symbols in Fig.~\ref{NegaPhase}
are inside the U phase (filled circles), T${}_{\theta<0}$ phase
(down triangles), T${}_{\theta>0}$ phase (up triangles), and S phase
(squares). We see that the actual U-T${}_{\theta<0}$ phase boundary
is well described by the variational method, but one can notice a
move of the S phase upwards and towards negative $\tilde\mu$. In
fact, the vacuum-S-T${}_{\theta>0}$ tri-critical point moves to
$\tilde\mu\approx-1.27$, $\gamma=0.78$ (outside of the plot).

In Fig.~\ref{NegaPhase} we also show density profiles corresponding
to the points enclosed by rectangular frames in the phase diagram.
The blue and yellow colors stand for minima and maxima of the
density. Without this rescaling the contrast, for instance, in the
lowest rightmost picture would be very weak. However, one can
clearly distinguish smooth density profiles, which can be described
by a few harmonics in the spirit of Eqs.~(\ref{superpsi1}) and
(\ref{psistripe}), and sharper profiles (as one moves further away
from the four-critical point) requiring more harmonics or a
different ansatz. The spatial coordinates have also been rescaled
(except for the upper set in the violet frame) because the wave
vector $k_m$ changes very strongly from point to point.

\begin{figure}
\includegraphics[width=1.\hsize,clip]{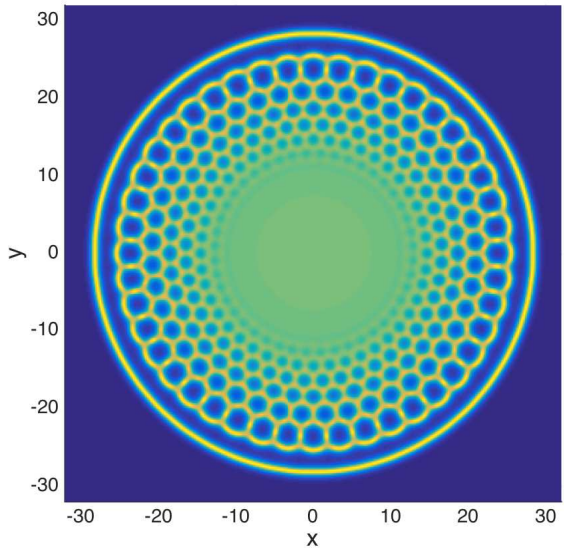}
\caption{(color online) The density profile for a harmonically
trapped gas with $\tilde\mu=0.6$, $\gamma=0.575$, and trapping
frequency $\tilde\omega=0.05$. The coordinates $x,y$ are in units of
$\sqrt{\hbar^2g_3/mg_2^2}$.}\label{trap}
\end{figure}

To the right of the vacuum curve (empty circles in
Fig.~\ref{NegaPhase}) the pressure is $P=-\Omega>0$ and, therefore,
this region of the phase diagram requires an external trapping. In
Fig.~\ref{trap} we present the exact GP result for an isotropically
trapped gas with $g_2<0$, $\gamma=0.575$, the global chemical
potential $\tilde\mu=0.6$, and trap frequency $\tilde\omega=0.05$
(in units of $g_2^2/\hbar g_3$). The result is consistent with the
local density approximation in which moving from the trap center
towards its edge is equivalent to the trajectory along a horizontal
line in Fig.~\ref{NegaPhase} determined by the local chemical
potential $\mu(r)= \mu -m \omega^2 r^2/2$. In Fig.~\ref{trap} one
can clearly distinguish the U phase in the trap center, the
transition to the T${}_{\theta<0}$ phase, and eventually to the S
phase. As the local chemical potential decreases, the contrast and
the period of the density modulation increase, which is consistent
with the free space results.

We should point out that first-order transitions involving density
jumps are forbidden in 2D systems with dipolar interaction tails.
The reason is that the surface tension in between two such phases
can have a negative contribution which logarithmically diverges with
the length of the interface and can thus overcome the positive local
scale-independent contribution \cite{Spivak} (see also
\cite{prokofiev}). This means that the first-order transition curves
that we describe here become (narrow) regions of intermediate
``microemulsion'' phases \cite{Spivak}. It is argued
\cite{prokofiev, Moroni} that the observation of these phases
requires exponentially large system sizes which are likely much
larger than the size of a typical ultracold sample. Nevertheless, we
note that already the simplest vacuum-U interface that we predict in
our dilute weakly-interacting system should be a good candidate for
studying these interfacial effects. However, we leave this subject
for future work.

In conclusion, we have found that a dilute 2D dipolar Bose gas can
reside in a variety of supersolid phases stabilized by three-body
repulsion. Our results represent a starting point for the analysis
of collective modes of homogeneous, trapped or self-trapped
supersolids.  The developed approach can also be employed in the
studies of novel vortex and soliton structures, and in the search
for translationally nonperiodic phases, in particular
density-disordered superfluid (superglass) phases. Promising
candidates for the creation of such dipolar Bose gases are
(nonreactive) polar molecules in the bilayer geometry with
interlayer tunneling. At 2D densities $n\sim 10^8$ cm$^{-2}$, for
the dipole moment $d\sim 0.5$ D one has $r_*\sim 200$ nm and
$nr_*^2\ll 1$. Then $g_3$ can be made such \cite{Petrov} that
$\gamma \sim 1$ and one may cover the whole range of parameters in
the phase diagram of Fig.~\ref{NegaPhase}. Finally, our results have
implications for magnetic atoms such as erbium or dysprosium, which
are necessarily dilute due to their small $r_*$. However, a
mechanism for generating a sufficiently strong three-body repulsion
in such gases has yet to be discussed.

We acknowledge support from IFRAF and from the Dutch Foundation FOM.
The research leading to these results has received funding from the
European Research Council under European Community's Seventh
Framework Programme (FR7/2007-2013 Grant Agreement no.341197). LY
acknowledges the support from the ARC Discovery Projects (Grant Nos
DE150101636 and DP140103231).

\newpage

\begin{widetext}

\section*{Supplementary Material}

In the Supplemental Material we first derive explicit expressions
for the energy functionals of the triangular and stripe supersolid
states, in particular the expressions for the functions
$\mathcal{C}_{T(S)}$, $\mathcal{D}_{T(S)}$, and
$\mathcal{T}_{T(S)}$. The second part of the Supplemental Material
is dedicated to the description of our numerical procedure of
minimizing the grand potential density.

\subsection{Derivation of the energy functionals for the triangular
and stripe supersolid states}

For calculating the energy functional we replace the field operators
$\hat\psi({\bf r})$ in the Hamiltonian (1) of the system with the
condensate wavefunction $\psi_T({\bf r})$ (7) for the triangular
supersolid and with $\psi_S({\bf r})$ (8) for the stripe phase. In
both cases the kinetic energy term proves to be
\begin{equation}     \label{K}
{\cal K}=-\int d^2r\psi^*_{T(S)}({\bf
r})\frac{\hbar^2\nabla^2}{2m}\psi_{T(S)}({\bf
r})=\frac{\hbar^2k^2n}{2m}\sin^2\theta.
\end{equation}

In the calculation of the contribution of the two-body interaction
we use the effective momentum-dependent interaction amplitude of
Eq.(2). Substituting this amplitude into equation (3) in which the
field operators $\hat\psi({\bf r})$ are replaced with the condensate
wavefunction $\psi_{T(S)}({\bf r})$ we obtain:
\begin{eqnarray}
{\cal H}_2^{T(S)}&=&{\cal H}_{2c}^{T(S)}+{\cal H}_{2d}^{T(S)},    \label{H2+} \\
{\cal H}_{2c}^{T(S)}&=&\frac{g_2}{2}\int d^2r|\psi_{T(S)}({\bf r})|^4, \label{H2c+} \\
{\cal H}_{2d}^{T(S)}&=&\frac{1}{2}\int d^2r |\psi_{T(S)}({\bf
r})|^2f(|{\bf r}-{\bf r}'|)|\psi_{T(S)}({\bf r}')|^2, \label{H2d+}
\end{eqnarray}
where Eqs. (\ref{H2c+}) and (\ref{H2d+}) represent the contributions
of the contact and dipole-dipole interactions, respectively, and the
function $f(|{\bf r}-{\bf r}'|)$ writes:
\begin{equation}     \label{f}
f(|{\bf r}-{\bf r}'|)=-\pi d^2 \int\frac{d^2q}{(2\pi)^2}q\exp(i{\bf
q}({\bf r}-{\bf r}')).
\end{equation}
The integration yields:
\begin{eqnarray}
{\cal H}_{2c}^{T(S)}&=&g_2n^2\mathcal{C}_{T(S)}(\theta), \label{H2c} \\
{\cal H}_{2d}^{T(S)}&=&-4\pi n^2d^2k\mathcal{D}_{T(S)}(\theta).
\label{H2d}
\end{eqnarray}
For the triangular phase the functions $\mathcal{C}_T(\theta)$ and
$\mathcal{D}_T(\theta)$ are given by
\begin{eqnarray}
\mathcal{C}_T(\theta)&=&\frac{1}{2}\left(\cos^4\theta +6\cos^2\theta\sin^2\!\theta +4\sqrt{\frac{2}{3}} \cos\theta\sin^3\!\theta+\frac{5}{2}\sin^4\theta\right),\label{CT} \\
\mathcal{D}_{T}(\theta)&=&\left[ \cos^2\!\theta
+\sqrt{\frac{2}{3}}\cos\theta\sin\theta+
\left(\frac{1}{4}+\frac{1}{2\sqrt{3}}\right)\sin^2\theta \right],
\label{DT}
\end{eqnarray}
and for the stripe phase we have:
\begin{eqnarray}
\mathcal{C}_{S}(\theta)&=&\frac{1}{2}\left(1+4\sin^2\theta\cos^2\theta+\frac{1}{2}\sin^4\theta\right),  \label{CS}   \\
\mathcal{D}_{S}(\theta)&=&\left(1-\frac{3}{4}\sin^2\theta\right).
\label{DS}
\end{eqnarray}

The integration of the third term of Eq.(1), representing the
contribution of the three-body contact interaction,
\begin{equation}    \label{H3+}
{\cal H}_3^{T(S)}=\frac{g_3}{6}\int d^2r|\psi_{T(S)}({\bf r})|^6,
\end{equation}
leads to
\begin{equation}      \label{H3}
{\cal H}_3^{T(S)}=g_3n^3\mathcal{T}_{T(S)}(\theta).
\end{equation}
The expressions for the functions $\mathcal{T}_T$ and
$\mathcal{T}_S$ read:
\begin{eqnarray}
 \mathcal{T}_{T}(\theta)&=&\frac{1}{6}\left(\cos^6\!\theta + 15\cos^4\!\theta\sin^2\!\theta  +20\sqrt{\frac{2}{3}}\cos^3\!\theta\sin^3\!\theta+\frac{75}{2}\cos^2\!\theta\sin^4\!\theta  +30\sqrt{\frac{2}{3}}\cos\theta\sin^5\!\theta+\frac{85}{9}\sin^6\!\theta\right). \label{TT} \\
\mathcal{T}_{S}(\theta)&=&\frac{1}{6}\left(1+12\sin^2\theta-
\frac{9}{2}\sin^4\theta-6\sin^6\theta\right). \label{TS}
\end{eqnarray}

The summation of ${\cal K}$ (\ref{K}), ${\cal H}^i_{2c}$
(\ref{H2c}), ${\cal H}^i_{2d}$ (\ref{H2d}), and ${\cal H}_3^i$
(\ref{H3}), where the symbol $i$ stands for $T$ and $S$, leads to
the energy functional in the form (9).

\subsection{Numerical procedure}

The numerical calculation is performed in the grand canonical
ensemble, with a given chemical potential $\mu$ and fixed volume of
the system $V = L_x L_y$. The field operator $\hat{\psi}$ is treated
as a classical field, and is discretized on a two-dimensional grid
with periodic boundary conditions in the coordinate and momentum
space. The grand potential reads:
\begin{equation}
\begin{aligned}
\Omega[\psi^\ast,\,\psi] &= \int d^2r\, \psi^\ast(\mathbf{r}) h_0
\psi(\mathbf{r})+ \frac{1}{2} \int d^2r\, d^2r'\, f(\mathbf{r} -
\mathbf{r}')|\psi(\mathbf{r}')|^2 |\psi(\mathbf{r})|^2 \\
& + \frac{g_2}{2} \int d^2r\, |\psi(\mathbf{r})|^4 + \frac{g_3}{6}
\int d^2r\, |\psi(\mathbf{r})|^6 - \mu \int d^2r\,
|\psi(\mathbf{r})|^2, \label{eq:grand_pot}
\end{aligned}
\end{equation}
where the single-particle Hamiltonian $h_0$ includes a possible
presence of the harmonic trapping potential:
\begin{equation}     \label{h0}
h_0=-\frac{\hbar^2}{2m}\nabla^2+\frac{1}{2}m\omega^2 r^2.
\end{equation}
The number of grid points that we use along each direction ranges
from 64 to 128 in the absence of the trapping potential and from 512
to 1024 in presence of an isotropic harmonic trap.

The ground state is determined by minimizing the grand potential
(\ref{eq:grand_pot}) with the use of the \emph{conjugate gradient}
algorithm \cite{Recipes}. An ingredient of this method is the line
minimization, that is in each iteration the wavefunction is changed
as
\begin{equation}
\psi_{i+1}(\mathbf{r}) = \psi_i(\mathbf{r}) +\lambda \Delta
\bar{\psi}_i(\mathbf{r}),
\end{equation}
where $\psi_i(\mathbf{r})$ is the wavefunction in a current step,
and $\lambda$ is a real parameter chosen to minimize
\eqref{eq:grand_pot} along the proposed direction
$\Delta\bar{\psi}_i(\mathbf{r})$. This procedure allows us to find
the global minimum encountered when moving downhill in
$\Omega[\psi^\ast,\, \psi]$ along a line. Consequently, it improves
the efficiency of the calculation. The direction along which to move
$\psi_i(\mathbf{r})$ is constructed as
\begin{equation}
\Delta\bar{\psi}_i(\mathbf{r}) = \Delta\psi_i(\mathbf{r})+
\frac{\int d^2r\, \Delta\psi_i^\ast(\mathbf{r}) \left[\Delta
\psi_i(\mathbf{r})- \Delta\psi_{i-1}(\mathbf{r}) \right]}{\int
d^2r\, |\Delta\psi_{i-1}(\mathbf{r})|^2} \Delta
\bar{\psi}_{i-1}(\mathbf{r}), \label{eq:dpsi_conj}
\end{equation}
in order to be \emph{conjugate} with respect to the direction
$\Delta\bar{\psi}_{i-1}(\mathbf{r})$ used in the previous step, and
\begin{equation}
\Delta\psi_i(\mathbf{r}) =  - \frac{\delta \Omega}{\delta\psi^\ast}
= - \left[\mathcal{H}_{\text{GP}}(\mathbf{r}) - \mu \right]
\psi_i(\mathbf{r}),
\end{equation}
is the gradient of the functional $\Omega[\psi^\ast,\,\psi]$
evaluated with $\psi_i(\mathbf{r})$, where
\begin{equation}
\mathcal{H}_{\text{GP}}  = h_0 + \int d^2r'\,
f(\mathbf{r}-\mathbf{r}') |\psi(\mathbf{r}')|^2 + g_2
|\psi(\mathbf{r})|^2 + \frac{g_3}{2} \,|\psi(\mathbf{r})|^4
\label{eq:H_GP}
\end{equation}
is the Gross-Pitaevskii Hamiltonian. The integral in the second term
of Eq.~(\ref{eq:H_GP}) can be calculated by using the convolution
theorem \cite{Goral2002}, namely
\begin{equation}
\int d^2r'\, f(\mathbf{r}-\mathbf{r}') |\psi(\mathbf{r}')|^2 =
\mathcal{F}^{-1} \left\{\mathcal{F} \left[f\right](\mathbf{q})\,
\mathcal{F} \left[|\psi|^2 \right](\mathbf{q}) \right\}
\end{equation}
where $\mathcal{F}\left[f\right](\mathbf{q})$ and $\mathcal{F}
\left[|\psi|^2 \right](\mathbf{q})$ are the Fourier transforms of
$f(\mathbf{r})$ and $|\psi(\mathbf{r})|^2$ respectively, and
$\mathcal{F}^{-1}$ is the inverse transform. We set that the
convergence is reached when the relative difference in the grand
potential between the neighboring time steps is smaller than
$5\times 10^{-9}$.

In the absence of external trapping, the wave function can remain
finite at the boundary. Due to the periodic boundary condition, the
structure of the modulation for a non-uniform state is then limited
by the size of the system imposed in the simulation. In order to
overcome this constraint, for each given set of parameters
$(g_2,\,g_3, \,d,\,\mu)$ we run the simulation several times with
different $L_x$ and $L_y$ ranging from $4\pi/k_m$ to $9 \pi/k_m$
respectively, where $k_m = 4\pi nr_\ast \mathcal{D}_{T(S)}(\theta)$
is fixed by the variational ansatz. In the end we choose the ground
state as the one corresponding to the lowest grand potential density
$\Omega / V$.

Different trial wavefunctions are used in the simulation, including
a uniform state, triangular (hexagonal) lattice state, square
lattice state, stripe state, a combination of triangular
(hexagonal) and stripe states. This is done in order to check
whether the final result is biased by the initial conditions or not.
We have also compared with each other the results obtained with a
different number of grid points to make sure that they are not
affected by the discretization of space.

\end{widetext}


\begin{thebibliography}{99}
\bibitem{note3} Here we focus on continuous space supersolids.
\bibitem{RemInfinitemu} In fact, one can think of the $g_2=0$ case as being represented by a vertical line in Fig.1 drawn at infinite $\tilde\mu$.
\end{thebibliography}
\end{document}